\documentclass[conference]{IEEEtran}
\IEEEoverridecommandlockouts
\usepackage{cite}
\usepackage{amsmath,amssymb,amsfonts}
\usepackage{algorithmic}
\usepackage{graphicx}
\usepackage{textcomp}
\usepackage{xcolor}
\usepackage{graphicx}
\usepackage{subfigure}
\usepackage{multicol}
\usepackage{setspace}
\usepackage{amsmath}
\usepackage{epstopdf}
\usepackage{fancyhdr}
\usepackage{indentfirst}
\usepackage{enumerate}
\usepackage{caption}
\usepackage{leftidx}
\usepackage{amssymb}
\usepackage{float}
\usepackage{breqn}
\usepackage{bbm}
\usepackage{booktabs}

\usepackage{multirow}
\usepackage{totcount}
\usepackage{algorithm}
\def\BibTeX{{\rm B\kern-.05em{\sc i\kern-.025em b}\kern-.08em
		T\kern-.1667em\lower.7ex\hbox{E}\kern-.125emX}}
\newtheorem{remark}{Remark}
\begin{document}
	\title{LEO Satellite-Enabled Grant-Free Random Access with MIMO-OTFS}
	
	\author{Boxiao Shen, Yongpeng Wu, Wenjun Zhang, Geoffrey Ye Li, Jianping An, and Chengwen Xing
	   	\thanks{The work of Y. Wu is supported in part by the National Key R\&D Program of China under Grant 2018YFB1801102, National Science Foundation (NSFC) under Grant 62122052 and 62071289, 111 project BP0719010, and STCSM 18DZ2270700.
	   	}
		\thanks{B. Shen, Y. Wu, and W. Zhang are with the Department of Electronic Engineering, Shanghai Jiao Tong
			University, Shanghai 200240, China (e-mail:
			boxiao.shen@sjtu.edu.cn; yongpeng.wu@sjtu.edu.cn; zhangwenjun@sjtu.edu.cn).
		}
		\thanks{G. Y. Li is with the Department of Electrical and Electronic
			Engineering, Imperial College London, London SW7 2AZ, UK (Email:
			geoffrey.li@imperial.ac.uk).}
		\thanks{J. An and C. Xing are with the School of Information and
			Electronics, Beijing Institute of Technology, Beijing 100081, China (e-mails: an@bit.edu.cn;  chengwenxing@ieee.org).} 
	}
	
	\maketitle
	
	\begin{abstract}
		This paper investigates joint channel estimation and device activity detection in the LEO satellite-enabled grant-free random access systems with large differential delay and Doppler shift. In addition, the multiple-input multiple-output (MIMO) with orthogonal time-frequency space modulation (OTFS) is utilized to combat the dynamics of the terrestrial-satellite link. To simplify the computation process, we estimate the channel tensor in parallel along the delay dimension. Then, the deep learning and expectation-maximization approach are integrated into the generalized approximate message passing with cross-correlation-based Gaussian prior to capture the channel sparsity in the delay-Doppler-angle domain and learn the hyperparameters. Finally, active devices are detected by computing energy of the estimated channel. Simulation results demonstrate that the proposed algorithms outperform conventional methods.
	\end{abstract}
	
	
	\section{Introduction}
	Internet-of-Things (IoT) are important application scenarios in the 5G and future networks. However, in many situations, a large number of IoT devices are distributed in remote areas, such as deserts, oceans, forests, etc\cite{s41}. The existing cellular communication networks are mainly designed and deployed in places where populations are concentrated, which makes it is difficult to support remote IoT devices. Fortunately, low-earth orbit (LEO) satellites are with low propagation delay, low path loss, and flexible elevation angle and could provide a very promising solution in these situations\cite{ss5}. Therefore, we are investigating the LEO satellite-enabled grant-free random access (RA) for IoT systems. 
	
	Over the past few years, many methods have been proposed for joint channel estimation and device activity detection in the terrestrial grant-free RA systems. For example, the sparse Bayesian learning (SBL) has been adopted in \cite{s36} to solve this problem. In \cite{s43}, the generalized multiple measurement vector approximate message passing (GMMV-AMP) has been proposed to explore the sparsity in the angular domain, which can further improve the estimation performance. The above works have been designed for the block fading channel, which is constant during a transmission slot. However, the high mobility of LEO satellites inevitably leads to the rapid change of terrestrial-satellite link (TSL) and the large Doppler shift, which degrades the performance of the current algorithms. Besides, the long delay effect should also be involved in the design of satellite communication systems. Therefore, the current grant-free random access schemes are hard to apply to LEO satellite communications straightforwardly. 
	
	Orthogonal time-frequency space (OTFS) which directly operates in the delay-Doppler domain is a promising solution to tackle the above issues. In \cite{s22} and \cite{s20}, the input-output relationship of OTFS with massive multiple-input multiple-output (MIMO) has been derived and the 3D-structured sparsity of the channel in the delay-Doppler-angle (DDA) domain has been identified, which enables the development of the 3D-structured orthogonal matching pursuit channel estimation algorithm. In \cite{smm1} and \cite{smm2}, two grant-free RA schemes with MIMO-OTFS have been proposed for the LEO satellite communications, where the device precompensates the delay and/or Doppler shift before the RA process. However, the precompensation incurs extra complexity, which will decrease the battery life of these remote IoT devices. 
	
	In this paper, we investigate the joint channel estimation and device activity detection in the LEO satellite-enabled grant-free RA, where OTFS is adopted to address the doubly dispersive effect in the TSL link and MIMO is integrated to improve the performance. Different from the existing literature, we consider the case that the differential delay is more than one symbol duration and/or the differential Doppler shift is more than one subcarrier spacing. To reduce the complexity of the IoT devices, we handle the large differential delay and Doppler shift at the satellite end. In this scenario, channel at each antenna can be expressed as a 3D tensor, which is complicated to handle directly. Hence, we estimate channel along delay dimension so that the 3D tensor can be decomposed into a set of matrices and computations can perform in parallel. Notice that the channel tensor in each delay dimension has the 2D burst block sparsity due to the adoption of MIMO-OTFS. To capture this 2D sparsity, we propose a deep learning-based generalized approximate message passing (DL-GAMP) algorithm with cross-correlation-based Gaussian prior, where the GAMP estimates the channel tensor, and the deep learning and expectation-maximization (EM) rule are adopted to learn the hyperparameters.

	\section{System Model}
	In this section, the basic system setup is first introduced. Then, we present the adopted channel model and input-output relationship of the MIMO-OTFS system when both the large differential delay and Doppler shift exist. Finally, we formulate the considered problem.
	\subsection{System Setup}
	We consider a LEO satellite-enabled grant-free random access system with MIMO-OTFS, where $U$ single-antenna devices adopting OTFS modulation intend to communicate with a LEO satellite. The satellite is equipped with $N_a = N_y \times N_z$ uniform planar array (UPA) antennas and works with the regenerative payload, which allows it for on-board processing of baseband signals. In a given time interval, only $U_a$ devices are active and transmit signal to satellite utilizing the same time-frequency resources. We denote $\lambda_u$ as the activity indicator of the $u$-th device, with $\lambda_u = 1$ if active and $\lambda_u = 0$ otherwise. We follow the recommendations of 3GPP\cite{3gpp} and assume that the satellite precompensates Doppler shift for the center of the beam and broadcasts a common delay to all devices. The residual delay and Doppler shift (called differential delay and Doppler shift in 3GPP terminology) are handled by satellite in the uplink transmissions. 
	\subsection{Channel Model and Signal Modulation}
	\subsubsection{Channel Model}We consider the following channel response corresponding to the $u$-th device in the delay-Doppler-space domain, i.e., 
	\begin{align}
	\label{cc}
	\mathbf h_u(\tau, \nu)=\sum_{i=0}^{P} h_{u,i} \delta\left(\tau-\tau_{u,i}\right) \delta\left(\nu-\nu_{u,i}\right) \mathbf v(\psi_{u,i},\varphi_{u,i}),
	\end{align} 
	where $h_{u,i}$, $\nu_{u,i}$, $\tau_{u,i}$, $\psi_{u,i}$, and $\varphi_{u,i}$ denote the complex gain, differential Doppler shift, differential delay, azimuth angle, and elevation angle of the $i$-th path, respectively, and $\mathbf v(\psi_{u,i},\varphi_{u,i})$ is the steering vector of the antennas. The steering vector is the Kronecker product of the array response vectors $\mathbf v(\vartheta_{y_{u,i}})$ and $\mathbf v(\vartheta_{z_{u,i}})$ corresponding to the directions with respect to $y$- and $z$-axis, respectively, that is,
	\begin{align}
	\mathbf v(\psi_{u,i},\varphi_{u,i}) = \mathbf v(\vartheta_{y_{u,i}}) \otimes \mathbf v(\vartheta_{z_{u,i}}),
	\end{align}
	where $\vartheta_{y_{u,i}} = \sin \varphi_{u,i} \sin \psi_{u,i}$ and $\vartheta_{z_{u,i}} = \cos \varphi_{u,i} $ are the directional cosines along the $y$- and $z$-axis, respectively, and
	\begin{align}
	\mathbf v(\vartheta) =\left[1 \exp \left\{-\bar{\jmath} \pi \vartheta\right\} \ldots\right.
	\left.\exp \left\{-\bar{\jmath} \pi\left(N_{y}-1\right) \vartheta\right\}\right]^{T}
	\end{align}
	
	The Doppler and delay taps for the $i$-th path of the $u$-th device can be expressed as
	\begin{align}
	\label{tap}
	\nu_{u,i} = \frac{k_{u,i} + \tilde{k}_{u,i} + b_{u,i}N}{NT_{\text{sym}}}, \tau_{u,i}=\frac{l_{u,i} + c_{u,i}M}{M\Delta f}, 
	\end{align}
	where $N$ is the number of OFDM symbols, $M$ is the number of subcarriers, and $\Delta f$ is the subcarrier spacing; $l_{u,i} = 0,\dots,M-1$ and $k_{u,i} = \lceil -N/2\rceil,\dots,\lceil N/2 \rceil -1$ represent indexes of the delay tap and Doppler tap corresponding to the delay $\tau_{u,i}$ and Doppler $\nu_{u,i}$, respectively; $\tilde{k}_i \in (-\frac{1}{2},\frac{1}{2}]$ corresponds to the fractional Doppler shift; $T_{\mathrm{sym}}$ is the time duration of a OFDM symbol with cyclic prefix (CP); $b_{u,i}$ and $c_{u,i}$ are integers, referred to as the outer Doppler tap and outer delay tap, respectively, which are non-zeros when differential Doppler $\nu_{u,i} > \Delta f$ and/or differential delay $\tau_{u,i} > T$ ($T$ is one symbol duration without CP). Outer Doppler and delay taps are often non-zeros in the LEO satellite communications due to the large geographical difference of devices, which is distinct from the terrestrial communications ($\nu_{u,i} \ll \Delta f$ and $\tau_{u,i} \ll T$). For example, in the 3GPP set-2 configuration\cite{3gpp}, the maximum differential Doppler and delay can be up to 3880 Hz and 699 $\mu$s\cite{ss5}, respectively. Then, if a subcarrier spacing is equal to 3880 Hz, a symbol duration (without CP) will be 258 $\mu$s and smaller than 699 $\mu$s. 
	\subsubsection{Signal Modulation}The $u$-th device employs the OFDM-based OTFS modulation to combat the doubly fading effect of the above channel. The detailed mathematical derivations of this modulation are omitted due to the limited spacing and interested readers can refer to \cite{s22,s20,smm1}. 
	We directly give the input-output relationship of the $u$-th device in the delay-Doppler-space domain at the $(n_z+n_yN_z)$-th antenna as\cite{smm1}
	\begin{align}
	&Y_{u, n_y,n_z}^{\mathrm{DDS}}[k, l] 
	=\sum_{i=0}^P \sum_{k^{\prime}=\lceil-N / 2\rceil}^{\lceil N / 2\rceil-1} X_u^{\mathrm{DD}}\left[\left\langle k-k^{\prime}\right\rangle_{N},\left(l-l_{u,i}\right)_{M}\right]\nonumber \\
	&\times  H_{u, n_y,n_z}^{\mathrm{DDS}}\left[k^{\prime}, l_{u,i}, l\right] + Z_{ n_{y}, n_{z}}^{\mathrm{DDS}}\left[k, l\right]
	\end{align}
	where $n_y = 0,\dots,N_y-1$, $n_z = 0,\dots,N_z-1$, $(\cdot)_{M}$ denotes mod $M$, $\langle x\rangle_{N}$ denotes $\left(x+\left\lfloor\frac{N}{2}\right\rfloor\right)_{N}-\left\lfloor\frac{N}{2}\right\rfloor$, $Z_{ n_{y}, n_{z}}^{\mathrm{DDS}}$ is the noise, and the effective channel $H_u^{\mathrm{DDS}}$. The effective channel can be expressed as
	\begin{align}
	\label{channelDDS}
	H_{u, n_y,n_z}^{\mathrm{DDS}}&\left[k^{\prime}, l_{u,i}, l\right] 
	= \frac{1}{N} \sum_{j=0}^{N-1} h_{M_{\mathrm{cp}}+j\left(M+M_{\mathrm{cp}}\right), l_{u,i}} e^{-\bar{\jmath} \frac{2 \pi}{N} k^{\prime} j}  \nonumber\\
	&\times e^{\bar{\jmath} 2 \pi \frac{\left(k_{u,i}+\tilde{k}_{u,i}+b_{u,i} N\right)\left(l-c_{u,i} M\right)}{\left(M+M_{\mathrm{cp}}\right) N}}e^{\bar{\jmath} \pi n_y \vartheta_{y_{u,i}}} e^{\bar{\jmath} \pi n_z \vartheta_{z_{u,i}}},
	\end{align}
	where $M_{\text{cp}}$ is the length of the CP and $h_{\rho, l}$ is the complex gain of the time-variant channel on the delay tap $l$ at the time $\rho T_{\mathrm{s}}$, defined as \cite{s20}
	\begin{align}
	h_{\rho, l}=\sum_{i=0}^{P} h_{{u,i}} e^{\bar{\jmath} 2 \pi(\rho-l) T_{s} \nu_{u,i}} \delta\left(l T_{\mathrm{s}}-(\tau_{u,i})_T\right),
	\end{align}
	where $T_{\mathrm{s}}=\frac{1}{M \Delta f}$ is the system sampling interval. 
	\begin{remark}
		The difference between the effective channel in (\ref{channelDDS}) and those in previous literature\cite{s22,s20,smm1} is that the phase compensation $e^{\bar{\jmath} 2 \pi \frac{\left(k_{u,i}+\tilde{k}_{u,i}+b_{u,i} N\right)\left(l-c_{u,i} M\right)}{\left(M+M_{\mathrm{cp}}\right) N}}$ is difficult to be separated from the effective channel or to be neglected, due to the existence of outer Doppler tap $b_i$ and outer delay tap $c_i$. In this scenario, the effective channel at each antenna is a 3D tensor and is complicated to estimate directly. To reduce computational complexity, we propose to handle it along delay dimension, subsequently.
	\end{remark}
	\subsection{Problem Formulation}
	To exploit the sparsity in the angular domain, the two-dimensional discrete Fourier transform (2D-DFT) is applied for the channel of $u$-th device along the space dimension, and then the delay-Doppler-angle domain channel is given by
	\begin{align}
	\label{angleD}
	&H_{u, a_y,a_z}^{\mathrm{DDA}}[k, l_{u,i}, l] 
	=\sqrt{N_yN_z}\sum_{i=0}^{P}h_{u,i} e^{\bar{\jmath} 2 \pi(M_{\text{cp}} + l)T_s\nu_{u,i}} \nonumber \\ 
	&\times e^{-\bar{\jmath} 2 \pi\tau_{u,i}\nu_{u,i}} \Pi_N(k-NT_{\text{sym}}\nu_{u,i})\delta\left(l_{u,i} T_{\mathrm{s}}-(\tau_{u,i})_T\right)\nonumber\\  &\times\Pi_{N_y}(a_y-N_y\vartheta_{y_{u,i}}/2) \Pi_{N_z}(a_z-N_z\vartheta_{z_{u,i}}/2),
	\end{align}
	where $a_y = 0,\dots,N_y-1$ and $a_z = 0,\dots,N_z-1$ are indexes along angular domain and  $\Pi_N(x)\triangleq\frac{1}{N} \sum_{i=0}^{N-1} e^{-\bar{\jmath} 2 \pi \frac{x}{N} i}$. From (\ref{angleD}), $H_{u, a_y,a_z}^{\mathrm{DDA}}[k, l_{u,i}, l]^{\mathrm{DDA}}$ has dominant elements only if $k \approx NT_{\text{sym}}(\nu_{u,i})_{1/T_{\text{sym}}}$, $l_{u,i} \approx M \Delta f(\tau_{u,i})_T$, $a_y \approx N_y\Omega_{y_{u}}/2$, and $a_z \approx N_z\Omega_{z_{u}}/2$. Therefore, the channel in the delay-Doppler-angle domain has the 3D-structured sparsity\cite{s22}. 
	
	The received symbols from all the devices in the delay-Doppler-angle domain is given by
	\begin{align}
	\label{ry}
	&Y_{a_y, a_z}^{\mathrm{DDA}}[k, l] =\sum_{u=0}^{U-1}\sum_{l^{\prime}=0}^{M-1} \sum_{k^{\prime}=\lceil-N / 2\rceil}^{\lceil N / 2\rceil-1} \lambda_u H_{u, a_y,a_z}^{\mathrm{DDA}}\left[k^{\prime}, l^{\prime}, l\right] \nonumber\\
	&\times X_u^{\mathrm{DD}}\left[\left\langle k-k^{\prime}\right\rangle_{N},\left(l-l^{\prime}\right)_{M}\right] + Z^{\mathrm{DDA}}[k, l,a_y, a_z],
	\end{align}  
	where $Z^{\mathrm{DDA}}$ is the noise in the delay-Doppler-angle domain. To facilitate the following analysis, we rewrite (\ref{ry}) into the matrix form for each $l$ as
	\begin{align}
	\label{incompletey}
	\mathbf Y_l =  \mathbf X_l\Lambda \tilde{\mathbf H}_l + \mathbf Z_l,
	\end{align}
	where the elements of $\mathbf Z_l$ are independent zero-mean Gaussian noises with variance $\sigma^2$; the $(a_z+1+N_za_y)$-th column of $\mathbf Y_l$ is $\mathbf Y_{a_y, a_z}^{\mathrm{DDA}}[:,l]$; 
	$\mathbf X_l=[\mathbf X_{0,l},\dots,\mathbf X_{U-1,l}]\in \mathrm{C}^{N\times UMN}$, where $\mathbf X_{u,l}\in\mathrm{C}^{N\times MN}$ is the pilot matrix of the $u$-th device in delay dimension $l$. Note that $\mathbf X_{u,l}$ is a block circulant matrix due to the 2D circular convolution in (\ref{ry}), 
	and its sub-matrix $\mathbf X_{u,l,l^{\prime}}$, $l^{\prime}=0,\dots,M-1$ is circulant matrix, formed by $\left[X_{u}^{\mathrm{DD}}[0,(l-l^{\prime})_M],\dots,X_{u}^{\mathrm{DD}}[-1,(l-l^{\prime})_M] \right]$. 
	$\Lambda = \mathrm{diag}([\lambda_0\mathbf{1}_{MN},\cdots,\lambda_{U-1}\mathbf{1}_{MN}])$ is a diagonal matrix composed of the activity indicator of all the devices and $\mathbf{1}_{MN} \in R^{MN}$ is a all-one vector; Finally, the $(a_z+1+N_za_y)$-th column of $\tilde{\mathbf H}$ is given by
	\begin{align}
	\label{H_vec}
	\tilde{\mathbf H}_l[:,a_z+1+N_za_y]=\left[\begin{array}{cccc}
	\mathrm{vec}\left(\mathbf H^{\mathrm{DDA}}_{0,a_y,a_z}[:,:,l]\right) \\
	\vdots \\
	\mathrm{vec}\left(\mathbf H^{\mathrm{DDA}}_{U-1,a_y,a_z}[:,:,l]\right)
	\end{array}\right],
	\end{align}
	where $\mathrm{vec}(\cdot)$ denotes the vectorization of a matrix.
	We denote the channel matrix $\Lambda \tilde{\mathbf H_l}$ as $\mathbf H_l$. Then, (\ref{incompletey}) can be written as
	\begin{align}
	\label{matrixform}
	\mathbf Y_l =\mathbf X_l\mathbf H_l + \mathbf Z_l.
	\end{align}
	Therefore, the joint device activity detection and channel estimation in this scenario turns into a sparse signal reconstruction problem, and can be separately and parallelly handled along delay dimension for each $\mathbf H_l$. In addition, the 2D burst block sparsity\cite{smm1} appears in $\mathbf H_l$, which is different from those in the conventional grant-free random access systems\cite{s36,s43} and can be utilized to improve the estimation performance further. 
	
	In this work, we estimate $\mathbf H_l$ adopting the minimum mean square error (MMSE) rule, represented as 
	\begin{align}
	\label{MMSE}
	\hat{\mathbf{H}}_l=\arg \max _{\mathbf{H^{\prime}}} \mathrm{E}[\|\mathbf{H^{\prime}}-\mathbf H_l\|_{2}^{2}|\mathbf Y_l]
	\end{align}
	Based on the estimation, the active device can be detected by comparing the energy of each device's channel with a pre-defined threshold, i.e.,
	\begin{align}
	\label{ddd}
	\hat{\lambda}_u = \mathbb{I}\left\{\frac{1}{M}\sum_{l=0}^{M-1}\sum_{i=uMN}^{(u+1)M N-1}\sum_{j=0}^{N_yN_z-1} \left|\hat H_{l,i,j}\right|^2>\xi_{th}\right\}, 
	\end{align}
	where $\hat H_{l,i,j}$ is the $(i,j)$-th element of $\hat{\mathbf{H}}_l$. Note that since each $\mathbf H_l$ can be estimated separately, we omit the index $l$ in the following content.
	\section{Joint Channel Estimation and Device\\ Activity Detection}
	\begin{figure*}[!htb]
		\centering
		\includegraphics[width=6in]{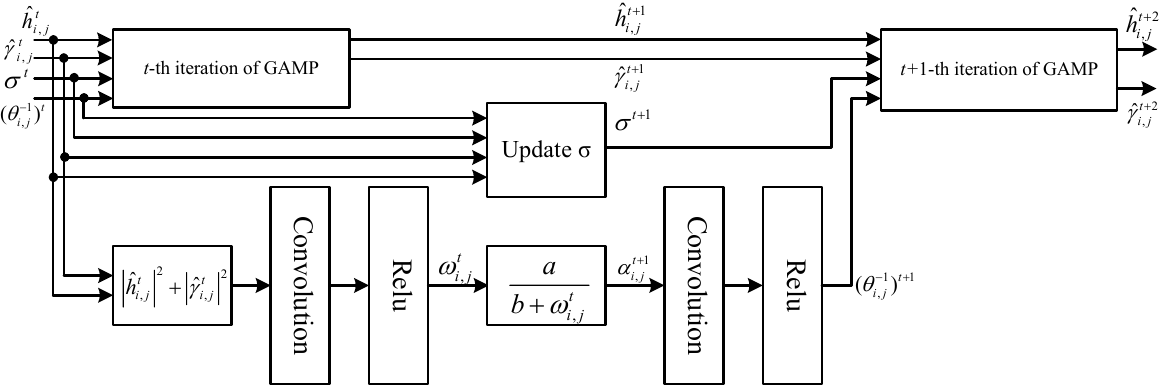}
		\caption{The proposed DL-GAMP algorithm.}
		\label{DLGAMP}
	\end{figure*}
	\subsection{Correlation-based Prior for Channel Coefficients}
	Since channel matrix $\mathbf H$ has the 2D burst block sparsity, i.e.,
	its non-zero elements are statistically dependent along the row and column, we consider the cross-correlation-based Gaussian prior for the $(i,j)$-th element of $\mathbf H$, given by
	\begin{align}
	\label{pcprior}
	p(h_{i,j}|\mathrm{\mathbf A},\mathrm{\mathbf B}) = \mathcal{CN}(h_{i,j}|0,\left(\sum_{p=-D_x}^{D_x}\sum_{q=-D_y}^{D_y} \beta_{p,q}\alpha_{i+p,j+q}\right)^{-1}),
	\end{align}
	where $\mathrm{\mathbf A}$ is the precision matrix whose $(i,j)$-th element $\alpha_{i,j}$ is the local precision hyperparameter of $h_{i,j}$; $\mathrm{\mathbf B}\in[0,1]^{D_x\times D_y}$ is the kernel and its  $(p,q)$-th element $\beta_{p,q}$ represents the relevence between $h_{i,j}$ and $h_{i+p,j+q}$. We denote $\theta_{i,j}$ as the prior variance of $h_{i,j}$.
	From (\ref{pcprior}), $\theta_{i,j}$  not only depends on local hyperparameter $\alpha_{i,j}$ of $h_{i,j}$, but also on the hyperparameters of its neighborhoods along the row and column of $\mathbf H$. Therefore,
	such coupled prior has the potential to capture the unknown 2D sparsity structure in $\mathbf H$. Then, following the conventional SBL, we adopt the Gamma distribution as the hierarchical prior for the local precision hyperparameter $\mathrm{\mathbf A}$, i.e.,
	\begin{align}
	p(\mathrm{\mathbf A}|a,b) = \prod_{i=0}^{UM N-1} \prod_{j=0}^{N_yN_z-1} \Gamma(a)^{-1} b^{a} \alpha_{i,j}^{a} e^{-b \alpha_{i,j}},
	\end{align}
	where $\Gamma(c)\triangleq\int_{0}^{\infty} t^{c-1} e^{-t} d t$. Note that this hierarchical Bayesian modeling encourages the sparseness in the estimation since the overall prior of $\mathbf H$ (i.e., integrate
	out $\alpha_{i,j}$) is a Student-$t$ distribution that is sharply peaked at zero. Based on the above prior distribution, it is possible to find the analytical solution of (\ref{MMSE}). However, the matrix inversion will inevitably be involved, which leads to the high computational complexity. Fortunately, the prior and the likelihood function of $\mathbf H$ can be fully factorized, leading to the following posterior distribution, i.e.,
	\begin{align}
	p( h_{i,j}&|\mathbf Y,\mathrm{\mathbf A},\mathrm{\mathbf B},\sigma^2) \propto \int p(\mathbf y_j|\mathbf g_j,\sigma^2)p(\mathbf h_j|\mathrm{\mathbf A}, \mathrm{\mathbf B}) dh_{\sim i,j}\nonumber\\
	= &\int\prod_{o=0}^{N-1}p(y_{o,j}|g_{o,j},\sigma^2) \prod_{i=0}^{UM N-1}p(h_{i,j}|\mathrm{\mathbf A},\mathrm{\mathbf B})dh_{\sim i,j},
	\end{align}
	where $\mathbf y_j$ and $\mathbf h_j$ are the $j$-th column of $\mathbf Y$ and $\mathbf H$, respectively, and $\mathbf g_j = \mathbf X\mathbf h_j$. Hence, we can resort to the GAMP algorithm to estimate the posterior mean and variance of the channel matrix with lower complexity.   
	\subsection{MMSE Estimation via GAMP} 
	The messages passed within the GAMP algorithm are approximated by Gaussian distributions, and then
	only the means and variances of the messages need to be calculated and preserved, which reduces the complexity significantly. 
	We next outline the MMSE estimation of channel matrix $\mathbf H$ in the $t$-th iteration by following the GAMP algorithms\cite{sss}. The detailed derivations are omitted for brevity. Here, the hyperparameters \{$\mathbf A$, $\mathbf B$, $\sigma$\} are assumed to be known and fixed, and we will update them later. 
	\subsubsection{Initilization ($t=1$)} The initial estimation of the posterior mean of the channel is set as its prior mean i.e., $\hat{h}_{i,j}^1=0$ and the estimation of the posterior variance $\hat \gamma_{i,j}$ is initialized to its prior variance $\theta_{i,j}$ with a small positive number, e.g., $\hat \gamma_{i,j}^1 = \theta_{i,j}^1 =0.01$. Residual $\hat{s}_{o,j}^0$ is set as 0.  
	\subsubsection{Messages on the Output Nodes} Based on linear relationship $\mathbf g_j = \mathbf X\mathbf h_j$ and observation $\mathbf y_j$, the messages are combined to obtain the posterior mean and variance of $g_{o,j}$, given by
	\begin{align}
	&\hat g_{o,j} = \mathrm E\left[g_{o,j} | y_{o,j}, \hat{p}_{o,j}^t,\tau_{o,j}^{p,t}\right], \nonumber\\
	&\tau_{o,j}^{g,t} = \mathrm{Var}\left[g_{o,j} | y_{o,j}, \hat{p}_{o,j}^t,\tau_{o,j}^{p,t}\right],
	\end{align}
	where $\mathrm E[\cdot]$ and $\mathrm{Var}[\cdot]$ are taken with respect to the posterior distribution of $g_{o,j}$ given prior $\mathcal{CN}(g_{o,j}|\hat{p}_{o,j}^t,\tau_{o,j}^{p,t})$ and likelihood function $\mathcal{CN}(y_{o,j}|g_{o,j},\sigma^2_t)$. The prior mean and variance of $g_{o,j}$ are calculated as 
	\begin{align}
	&\tau_{o,j}^{p,t}=\sum_{i}\left|X_{o,i}\right|^2 \hat \gamma_{i,j}^t, \nonumber \\
	&\hat{p}_{o,j}^t = \sum_{i} (X_{o,i}\hat{h}_{i,j}^t) -
	\hat{s}_{o,j}^{t-1}\tau_{o,j}^{p,t} \label{pp}.
	\end{align}
	Then, residual $\hat{s}_{o,j}^t$ and inverse-residual variance $\tau_{o,j}^{s,t}$ are computed by
	\begin{align}
	\label{ss1}
	&\hat{s}_{o,j}^t =\left(\hat g_{o,j}^t-\hat{p}_{o,j}^t\right) / \tau_{o,j}^{p,t}, \nonumber\\
	&\tau_{o,j}^{s,t} =\left(1-\tau_{o,j}^{g,t} / \tau_{o,j}^{p,t}\right) / \tau_{o,j}^{p,t}.
	\end{align}
	\subsubsection{Messages on the Input Nodes}
	Based on the residual and the inverse-residual variance, the posterior mean and variance of $h_{i,j}$ can be updated by
	\begin{align}
	\label{eh}
	&\hat h_{i,j}^{t+1} =  \mathrm{E}[h_{i,j}|\hat{r}_{i,j}^t,\tau^{r,t}_{i,j},\mathrm{\mathbf A^t},\mathrm{\mathbf B^t}], \nonumber\\
	&\hat\gamma_{i,j}^{t+1}=\mathrm{Var}[h_{i,j}|\hat{r}_{i,j}^t,\tau^{r,t}_{i,j},\mathrm{\mathbf A^t},\mathrm{\mathbf B^t}],
	\end{align}
	where $\mathrm E[\cdot]$ and $\mathrm{Var}[\cdot]$ are taken with respect to the posterior distribution of $h_{i,j}$ given prior $p(h_{i,j}|\mathbf A^t, \mathbf B^t)$ and likelihood function $\mathcal{CN}(h_{i,j}|\hat{r}_{i,j}^t,\tau^{r,t}_{i,j})$. The  mean and variance in the likelihood function are calculated as
	\begin{align}
	&\tau^{r,t}_{i,j}=\left(\sum_{o} \left|X_{o,i} \right|^2 \tau_{o,j}^{s,t}\right)^{-1}, \nonumber\\
	&\hat{r}_{i,j}^t=\hat h_{i,j}^{t}+\tau^{r,t}_{i,j} \sum_{o} X_{o, i}^{*} \hat{s}_{o,j}^t. \label{ee1}
	\end{align}
	Combining (\ref{pcprior}) and (\ref{eh})-(\ref{ee1}), we have
	\begin{align}
	&\hat h_{i,j}^{t+1} = \frac{\theta_{i,j}^t\hat{r}_{i,j}^t}{\theta_{i,j}^t+\tau^{r,t}_{i,j}}, \nonumber\\
	&\hat \gamma_{i,j}^{t+1} = \frac{\theta_{i,j}^t\tau^{r,t}_{i,j}}{\theta_{i,j}^t+\tau^{r,t}_{i,j}}.\label{mmm2}
	\end{align}
	\subsection{Learning the Hyperparameters}
	In this subsection, we utilize the EM algorithm together with deep learning to learn the hyperparameters \{$\mathbf A$, $\mathbf B$, $\sigma$\}, where $\mathbf A$ and $\sigma$ are updated by the EM-based rule and $\mathbf B$ is learned by a convolutional neural network (CNN), as shown in Fig. \ref{DLGAMP}. We then explain it in detail. Following the conventional SBL, we place a Gamma hyperprior over $\sigma^{-2}$, i.e.,
	\begin{align}
	p(\sigma|r,s) = \Gamma(r)^{-1} s^{r} \sigma^{-2r} e^{-s \sigma^{-2}}.
	\end{align}
	Based on the estimation in the $t$-th iteration, the precision matrix and the noise variance can be updated by
	\begin{align}
	&\mathrm{\mathbf A}^{t+1}=\arg \max _{\mathrm{A}} \mathrm{E}[\log p(\mathbf H|\mathrm{\mathbf A},\mathrm{\mathbf B})p(\mathrm{\mathbf A})|\mathbf Y, \mathbf A^{t}, \sigma^t]\label{Am},\\
	&\sigma^{t+1} = \arg \max _{\sigma} \mathrm{E}[\log p(\mathbf Y|\mathbf H,\sigma)p(\sigma)|\mathbf Y, \mathbf A^{t}, \sigma^t]\label{noisep},
	\end{align}
	where the kernel $\mathrm{\mathbf B}$ is assumed to be known in this step. 
	\addtolength{\topmargin}{0.03in}
	The optimization in (\ref{Am}) is difficult to get the analytic solution since the hyperparameters are entangled with each other due to the logarithm term. Although the gradient descent
	methods can be adopted to get the optimal solution, 
	it involves higher computational complexity as compared with an analytical update rule. We derive a closed-form sub-optimal solution (see \cite{smm1} for more detail) by examining the first derivative of the objective function with respect to $\alpha_{i,j}$ and by proper scaling, given as
	\begin{align}
	\label{alpha}
	\alpha_{i,j}^{t+1}= \frac{a}{b+\omega_{i,j}^{t}},
	\end{align}
	where $\omega_{i,j}^{t}=\sum_{p=-D_x}^{D_x}\sum_{q=-D_y}^{D_y}\beta_{p,q}\mathrm{E}[\left|h_{i-p, j-q}\right|^{2}]$ and posterior second moment $\mathrm{E}[\left|h_{p q}\right|^{2}]=\left|\hat h^t_{p,q}\right|^2 + \left|\hat \gamma^t_{p,q}\right|^2$, which are the outputs of the GAMP algorithm in the $t$-th iteration.
	Then, computing the first derivative of the objective function in (\ref{noisep}) with respect to $\sigma^2$ and setting it equal to zero, we get
	\begin{align}
	\label{theta_up}
	(\sigma^2)^{t+1}= \frac{\sum_{j=0}^{N_yN_z-1}\mathrm{E}\left[\left\|\mathbf{y}_j-\mathbf X\mathbf{h}_j\right\|^2\right] +s}{M N N_yN_z+r},
	\end{align}
	where
	\begin{align}
	\mathrm{E}\left[\left\|\mathbf{y}_j-\mathbf X\mathbf{h}_j\right\|^2\right] = &\left\|\mathbf{y}_j-\mathbf X \mathbf{\hat{h}}^t_j\right\|^2 
	+ (\sigma^2)^{t}\nonumber\\
	&\times\sum_{i=0}^{UM N-1}(1-\hat \gamma^t_{i,j}/\theta^t_{i,j}).
	\end{align}
	By examining (\ref{pcprior}) and (\ref{alpha}), the update of precision matrix $\mathbf A$ is related to the 2D convolution between kernel $\mathbf{B}$ and the posterior second moment, and the update of prior variance $\theta_{i,j}$ is related to the cross-correlation between the kernel and precision matrix, i.e., 
	\begin{align}
	\label{tup}
	(\theta_{i,j}^{-1})^{t+1} =\sum_{p=-D_x}^{D_x}\sum_{q=-D_y}^{D_y} \beta_{p,q}\alpha_{i+p,j+q}^{t+1}.
	\end{align}
	The only difference between the 2D convolution and the cross-correlation is whether the kernel is flipped within the computations, which motivates us to utilize the CNN to learn kernel $\mathbf B$ and then update precision matrix $\mathbf A$. Here, the trainable filter of the convolutional layer can be seen as flipped kernel $\mathbf B$. Combining this idea with the EM update rule and the GAMP algorithm leads to the deep learning-based GAMP algorithm (DL-GAMP), and we summarize the computation process as follows.
	
	As shown in Fig. \ref{DLGAMP}, in the $t$-th iteration, the posterior second moment computed by the GAMP algorithm in the $(t-1)$-th iteration is firstly inputted to a convolutional layer and then the rectified linear unit (ReLU)\cite{relu} will ensure a positive output $\omega_{i,j}^{t}$. Next, precision $\alpha_{i,j}^{t+1}$ is updated according to (\ref{alpha}), and then is inputted to another convolutional layer followed by the ReLU to get the update of prior variance $\theta_{i,j}^{t+1}$ (similar to (\ref{tup})). Simultaneously, noise standard variance $\sigma^{t+1}$ is given by (\ref{theta_up}). Finally, the output of GAMP in this iteration \{$\hat h_{i,j}^{t+1}$, $\hat \gamma_{i,j}^{t+1}$\} and the update of \{$\theta_{i,j}^{t+1}$, $\sigma^{t+1}$\} computed by the CNN and the EM approach will be inputted to the next iteration of the DL-GAMP. The algorithm will run until the maximum iteration $T_{\text{max}}$ is reached, and then the active device is detected according to (\ref{ddd}). Note that the filters in the two convolutional layers are not restricted to be identical since they will be learned from the training samples. Besides, the matrix multiplication in the GAMP algorithm dominates other computations, like the 2D convolution in CNN, leading to complexity $\mathcal{O}(UMN^2N_yN_z)$ in each iteration when each channel matrix $\mathbf H_l$ is parallelly estimated by DL-GAMP. If $\mathbf H_l$ is estimated in a serial manner, the complexity will be $\mathcal{O}(UM^2N^2N_yN_z)$ in each iteration.
	
	\section{Numerical Results}
	\begin{table}
		\scriptsize
		\caption{SIMULATION PARAMETERS}
		\begin{center}
			\begin{tabular}{|c|c|}
				\hline Parameter & Values \\
				\hline Carrier frequency $(\mathrm{GHz})$ & 2 \\
				\hline Subcarrier spacing $(\mathrm{kHz})$ & 3.75 \\
				\hline Size of OTFS symbol $(M, N)$ & $(16,35)$ \\
				\hline Length of CP  & 42 \\
				\hline Number of Satellite antennas & $4\times4$ \\
				\hline Number of LoS paths & 1 \\
				\hline Number of Non-LoS path & 1 \\
				\hline The differential delay (\textmu s) & $[0,699]$\\
				\hline The differential Doppler shift (kHz) & $[-3.88, 3.88]$\\
				\hline Directional cosine along the y- and z-axis & $[-1,1]$\\
				\hline Rician factor (dB) & $8$
				\\
				\hline
			\end{tabular}
			\label{spa}
		\end{center}
	\end{table}
	In this section, we demonstrate the performance of the proposed algorithms through computer simulations. We consider the set-2 configuration of the non-terrestrial networks recommended by the 3GPP\cite{ss5}\cite{3gpp}. The detailed system parameters are summarized in Table \ref{spa}. Each device transmits Gaussian pliot matrix with independent entries obeying $\mathcal{CN}(0,\frac{1}{M N})$. Moreover, the power of the multi-path channel gain
	is normalized as 1, i.e., $\sum_{i=0}^{p}\left|h_{u,i}\right|^2=1$. Finally, we define the signal-to-noise ratio (SNR) as $\text{ SNR }=10\log_{10}\frac{1}{MN\sigma^2}$. 
	
	To train the DL-GAMP, we generate $2\times10^{4}$ samples; set $D_x=D_y=1$, $b=r=s=10^{-4}$, and $T_{max}=600$; set $a$ as the hyerparameter learned in the training process; adopt the Adam optimizer to update parameters in CNN with the
	learning rate of $10^{-4}$. We generate $3\times10^{3}$ samples at each SNR point to test the DL-GAMP. Besides, we adopt GMMV-AMP\cite{s43} as benchmarks to show the effectiveness of the proposed algorithms. Finally, the average normalized mean-squared-error (NMSE) and the average error probability are adopted as the metrics for the channel estimation and device activity detection, respectively, given by
	\begin{align}
	\text{ NMSE }=\frac{1}{M}\sum_{l=0}^{M-1}\frac{\left\| \mathbf H_l-\hat{\mathbf H}_l\right\|^2_{\mathrm{F}}}{\left\|\mathbf H_l\right\|^2_{\mathrm{F}}},
	P_e = \frac{1}{U}\sum_{u=0}^{U-1}\left|\lambda_u-\hat \lambda_u\right|.
	\end{align} 
	
	\begin{figure}[!htb]
		\centering
		\includegraphics[width=2.6in]{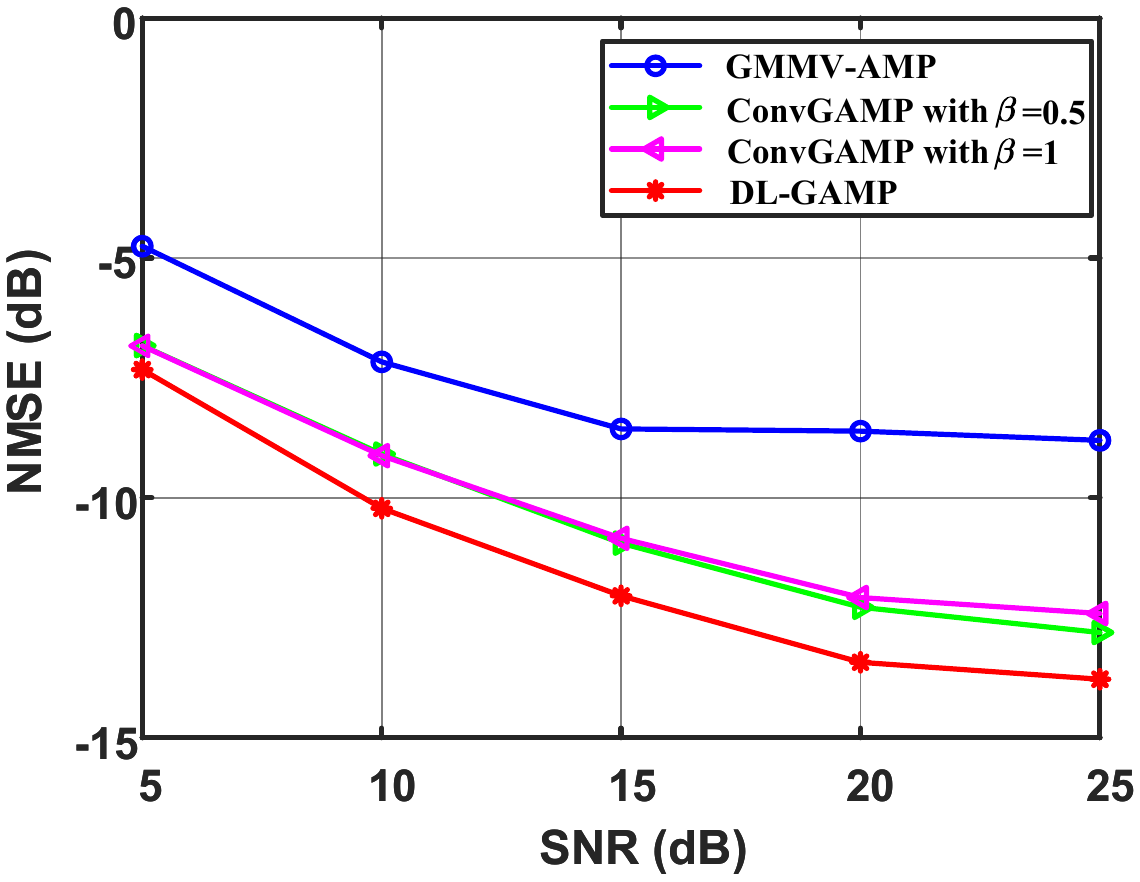}
		\caption{Performance comparison of channel estimation between the GMMV-AMP, ConvGAMP, and DL-GAMP under different SNR values. $U = 5$, $U_a=1$, $N_y=N_z=4$.}
		\label{NMSEF}
	\end{figure}

	\begin{figure}[!htb]
		\centering
		\includegraphics[width=2.6in]{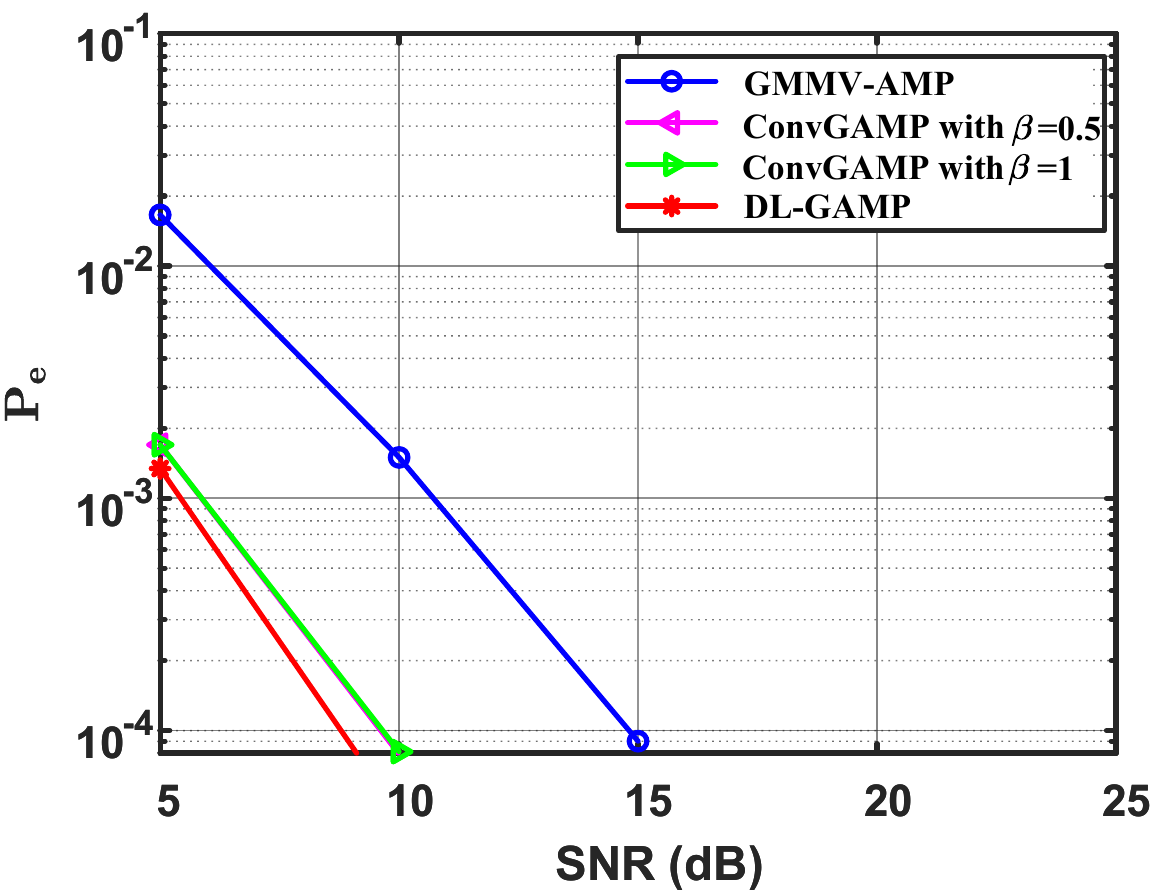}
		\caption{Performance comparison of device activity detection between the GMMV-AMP, ConvGAMP, and DL-GAMP under different SNR values. $U = 5$, $U_a=1$, $N_y=N_z=4$.}
		\label{PdF}
	\end{figure} 
	Fig. \ref{NMSEF} compares the  channel estimation performance between the DL-GAMP and GMMV-AMP. In the figure, we also provide the performance of the proposed algorithm with a constant kernel (without learning), named ConvGAMP, where the center coefficient of the kernel is set as 1 and other coefficients are set to be identical with $\beta = 0.5$ or $\beta = 1$, which means that the neighborhoods of the channel $h_{i,j}$ have the same influence on it. Besides, there are five potential devices with an active one, i.e., $U=5$ and $U_a=1$. From the figure, the proposed algorithms ConvGAMP and DL-GAMP outperform the GMMV-AMP under different SNR values since the GMMV-AMP only captures the sparsity in the angular domain while the proposed algorithms are able to deal with the sparsity in the delay-Doppler-angle domain.
	For example, when the NMSE is about -7.5 dB, the proposed algorithms outperform the GMMV-AMP by around 5 dB. Besides,  the DL-GAMP has better performance than ConvGAMP, and has the 2.5 dB gain when the NMSE is -10 dB, which indicates that the kernel learned by CNN can provide better estimation performance. 
	
	Finally, in Fig. \ref{PdF}, we compare the device activity detection performance of the GMMV-AMP, ConvGAMP, and DL-GAMP, where $U=5$ and $U_a=1$. From the figure, similar to Fig. \ref{NMSEF}, the ConvGAMP and DL-GAMP outperform the GMMV-AMP. With the assistance of CNN, the performance of DL-GAMP is improved further compared with the ConvGAMP.
	\section{Conclusion}
	This work investigated the application of MIMO-OTFS for the
	grant-free random access in LEO satellite
	communications, where there are usually large differential delay and Doppler shift. To exploit
	the 2D burst block sparsity in the delay-Doppler-angle
	domain, the correlation-based Gaussian prior with the GAMP was proposed to estimate the channel and detect the active device. Finally, the hyperparameters were learned by the EM rule and deep learning. Simulation results demonstrated that the proposed algorithms could acquire accurate channel state information and device activity.

	\bibliographystyle{IEEEtran}%
	\bibliography{bibfile}
	\vspace{12pt}

\end{document}